\documentclass[amsmath,amssymb,prl,superscriptaddress,preprint]{revtex4-1}

\usepackage{amsmath,amssymb,amsfonts,mathrsfs}
\usepackage{graphicx,soul}
\usepackage{color,ulem}
\usepackage{sans}
\usepackage{SIunits}

\newcommand{\fcc}{Fe$^{\rm{fcc}}$~}
\newcommand{\hcp}{Fe$^{\rm{hcp}}$~}

\newcommand {\jw}[1]{{\color{black} #1}}

\newcommand{\Feonefcc}{FeH$^{\rm{fcc}}$~}
\newcommand{\Fetwofcc}{FeH$_{2}^{\rm{fcc}}$~}
\newcommand{\Feonehcp}{FeH$^{\rm{hcp}}$~}
\newcommand{\Fetwohcp}{FeH$_{2}^{\rm{hcp}}$~}
\renewcommand{\Im}{\ensuremath{{\rm Im}\,}}

\begin{document}

\title{Tuning emergent magnetism in a Hund's impurity}

\author{A. A. Khajetoorians}
\email{Corresponding author: a.khajetoorians@science.ru.nl}
\affiliation{Institute of Applied Physics, Hamburg University, D-20355 Hamburg, Germany}
\affiliation{Institute for Molecules and Materials (IMM), Radboud University, 6525 AJ Nijmegen, The Netherlands.}

\author{M. Valentyuk}
\affiliation{I. Institute of Theoretical Physics, Hamburg University, D-20355 Hamburg, Germany}
\affiliation{Department of Theoretical Physics and Applied Mathematics, Ural Federal University, 620002 Ekaterinburg, Russia}

\author{M. Steinbrecher}
\affiliation{Institute of Applied Physics, Hamburg University, D-20355 Hamburg, Germany}

\author{T. Schlenk}
\affiliation{Institute of Applied Physics, Hamburg University, D-20355 Hamburg, Germany}

\author{A. Shick}
\affiliation{Institute of Physics, ASCR, Na Slovance 2, CZ-18221 Prague, Czech Republic}

\author{J. Kolorenc}
\affiliation{Institute of Physics, ASCR, Na Slovance 2, CZ-18221 Prague, Czech Republic}

\author{A. I. Lichtenstein}
\affiliation{I. Institute of Theoretical Physics, Hamburg University, D-20355 Hamburg, Germany}

\author{T. O. Wehling}
\affiliation{Institute for Theoretical Physics, Bremen Center for Computational Material Science, University of Bremen, D-28359 Bremen, Germany}

\author{R. Wiesendanger}
\affiliation{Institute of Applied Physics, Hamburg University, D-20355 Hamburg, Germany}

\author{J. Wiebe}
\email{jwiebe@physnet.uni-hamburg.de}
\affiliation{Institute of Applied Physics, Hamburg University, D-20355 Hamburg, Germany}



\begin{abstract}
\textbf{The recently proposed theoretical concept of a Hund's metal is regarded as a key to explain the exotic magnetic and electronic behavior occuring in the strongly correlated electron systems of multiorbital metallic materials. However, a tuning of the abundance of parameters, that determine these systems, is experimentally challenging. Here, we investigate the smallest possible realization of a Hund's metal, a {\it Hund's impurity}, realized by a single magnetic impurity strongly hybridized to a metallic substrate. We experimentally control all relevant parameters including magnetic anisotropy and hybridization by hydrogenation with the tip of a scanning tunneling microscope and thereby tune it through a regime from emergent magnetic moments into a multi-orbital Kondo state. Our comparison of the measured temperature and magnetic field dependent spectral functions to advanced many-body theories will give relevant input for their application to non-Fermi liquid transport, complex magnetic order, or unconventional superconductivity.}

\end{abstract}

\maketitle%





Recent examples of exotic phases of matter, including unconventional superconductivity in iron pnictides and chalcogenides~\cite{Stewart:2011, Yin:2011, Werner2012} as well as non-Fermi liquid behavior in ruthenates \cite{Werner2008,Mravlje2011,deMedici2011}, depend subtly on the complex interplay of magnetic moments and delocalized electron states taking place in transition metal $d$-shells. All these materials combine sizable Coulomb interactions and hybridization, which are comparable in their strength. In such cases, it is generally unclear, to which extent local magnetic moments exist, how they can be described using quantum impurity models~\cite{Anderson1961}, and how far electronic correlation effects such as Kondo screening~\cite{Kondo:1964,Nozieres:1980} modify material properties, particularly magnetism, as a function of temperature and magnetic field. The recent concept of a \textit{Hund's metal}~\cite{Kaule:2009, Yin:2011, Georges2013} has been introduced in order to describe exactly this regime, where charge fluctuations in the orbitals are not negligible due to the presence of strong hybridization, but where local magnetic moments can still survive.

\begin{figure}[t!]
\centerline{\includegraphics[width = \columnwidth]{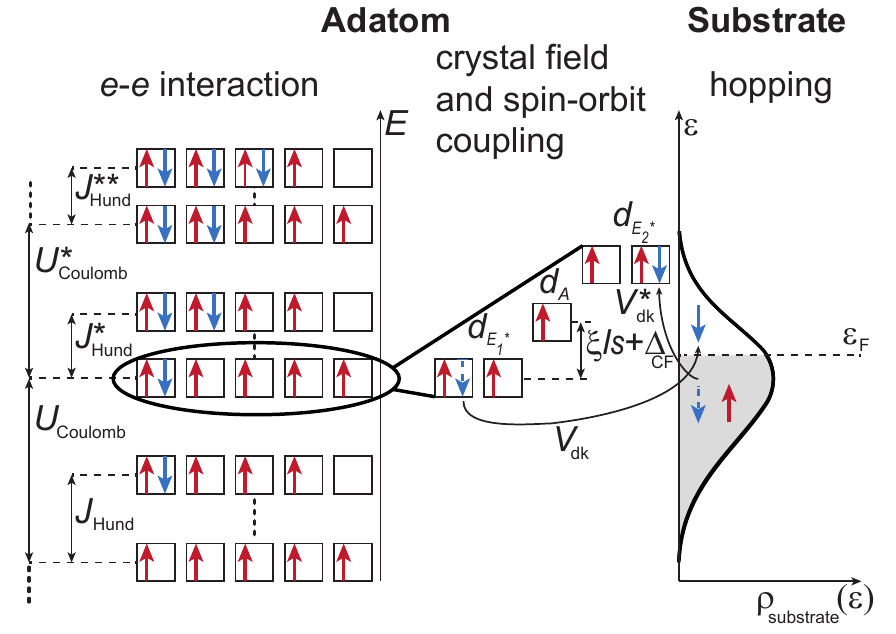}}
\caption{\label{fig1} {\bf Electrons in a 3$d$ transition metal adatom adsorbed to a metallic substrate.} Left panel: electrons of spin up (red arrows) or down (blue arrows) are filled into the five 3$d$ orbitals (boxes) of the adatom. Due to the Coulombic electron-electron ($e-e$) interaction, on-site Coulomb energy ($U_{\rm{Coulomb}}$) or Hund's rule exchange energy ($J_{\rm{Hund}}$) have to be paid if an additional electron is put into an orbital, or if one of the electron spins is flipped, respectively. Middle panel: Blow-up of the energetic positions of the five orbitals in one of the electron configurations, which are split by crystal field ($\Delta_{\rm{CF}}$) and spin-orbit coupling ($\xi ls $). Right panel: Hybridization ($V_{\rm{dk}}$) of the adatom orbitals and the substrate electron density of states $\rho_{\rm{substrate}}$ leads to hopping of electrons on and off of the impurity, and to charge fluctuations and non-(half) integer magnetic moments.}
\end{figure}

The fundamental constituent of such a Hund's metal is a magnetic impurity strongly coupled to the electron states of a metallic host, which we coin {\it Hund's impurity}.
This concept is described in the following for the particular case of a 3\textit{d} transition metal atom that gets adsorbed (adatom) onto a metallic substrate (Fig.~\ref{fig1}).
If the atom is still in the gas phase an integer number of electrons is filled into the five 3\textit{d} orbitals according to Hund's \jw{first} rule:~\cite{Hund:1927a,Hund:1927b} The orbitals are first filled up by electrons having the same spin, before being filled with the remaining electrons of opposite spin. This is driven by the intraatomic exchange energy, or so-called Hund's rule exchange $J_{\rm{Hund}}$, which has to be paid if one of the electron spins is flipped. If the 3\textit{d} transition metal atom is \jw{adsorbed onto the metallic substrate, electrons can hop on or off of these orbitals into the bath of substrate conduction electrons, which has an electron density of states $\rho_{\rm{substrate}}$,} paying or gaining on-site Coulomb energy $U_{\rm{Coulomb}}$. \jw{This hopping leads to} fluctuations of the charge in the orbitals. The strength of the hopping $V_{\rm{dk}}$ from the adatom to the bath~\cite{Carbone2010}, and the valency~\cite{Gardonio2013} dictate whether the electronic structure of the adatom can still be described by an atomic multiplet structure, itinerant electrons, or a degree of both with distinct correlation effects. For negligible hybridization $V_{\rm{dk}}\approx0$, \jw{referred to as an atomic ($\mathcal{A}$) impurity in the following,} the system can be understood in terms of \jw{crystal field splittings $\Delta_{\rm{CF}}$ and spin-orbit coupling (SOC) $\xi ls $ with a} well-defined valency and (half-) integer quantized spin~\cite{Heinrich2004, Hirjibehedin2007,Delgado2011,Lorente2012}. With weak hybridization, the adatom retains its integer valency but correlations between the atomic spin and the surface electrons, such as a Kondo singlet formation, can set in~\cite{Otte2008,Oberg2014}.
\jw{With even further increased hybridization, and if the adatom was formed solely by a single orbital, the adatom magnetic moment would be simply quenched at $V_{\rm dk} > \sqrt{\frac{U_{\rm{Coulomb}}}{\rho_{\rm{substrate}}(E_{\rm{F}})}}$ ($E_{\rm{F}}$ is the Fermi energy)~\cite{Anderson1961}}. However, for the multi-orbital case, and in a particular regime with strong hybridization which is yet too weak to overcome $J_{\rm{Hund}}$, strong charge fluctuations can coexist with sizeable local magnetic moments, which are strongly coupled to their environment. This is referred to as the Hund's impurity ($\mathcal{HI}$) regime~\cite{TW_Werner}. It is characterized by a complex interplay of charge fluctuations, $\Delta_{\rm{CF}}$, $\xi ls $, and electron correlations. The investigation of this $\mathcal{HI}$ regime is a current challenge for advanced theoretical methods~\cite{RMP_CTQMC, vDelft13}. Moreover, the experimental realization of a $\mathcal{HI}$, and, more importantly, the full control over all the relevant parameters, i.e. magnetic anisotropy, hybridization, temperature and magnetic field, remained incomplete so far~\cite{Madhavan1998, Nagaoka2002, Otte2008, Parks:2010, Khajetoorians2011, Mugarza:2011, Oberg2014,Donati2013,Serrate2014}.



\subsection{Manipulation of the magnetic properties of iron adatoms}
\begin{figure}[t!]
\centerline{\includegraphics[width = 0.75\columnwidth]{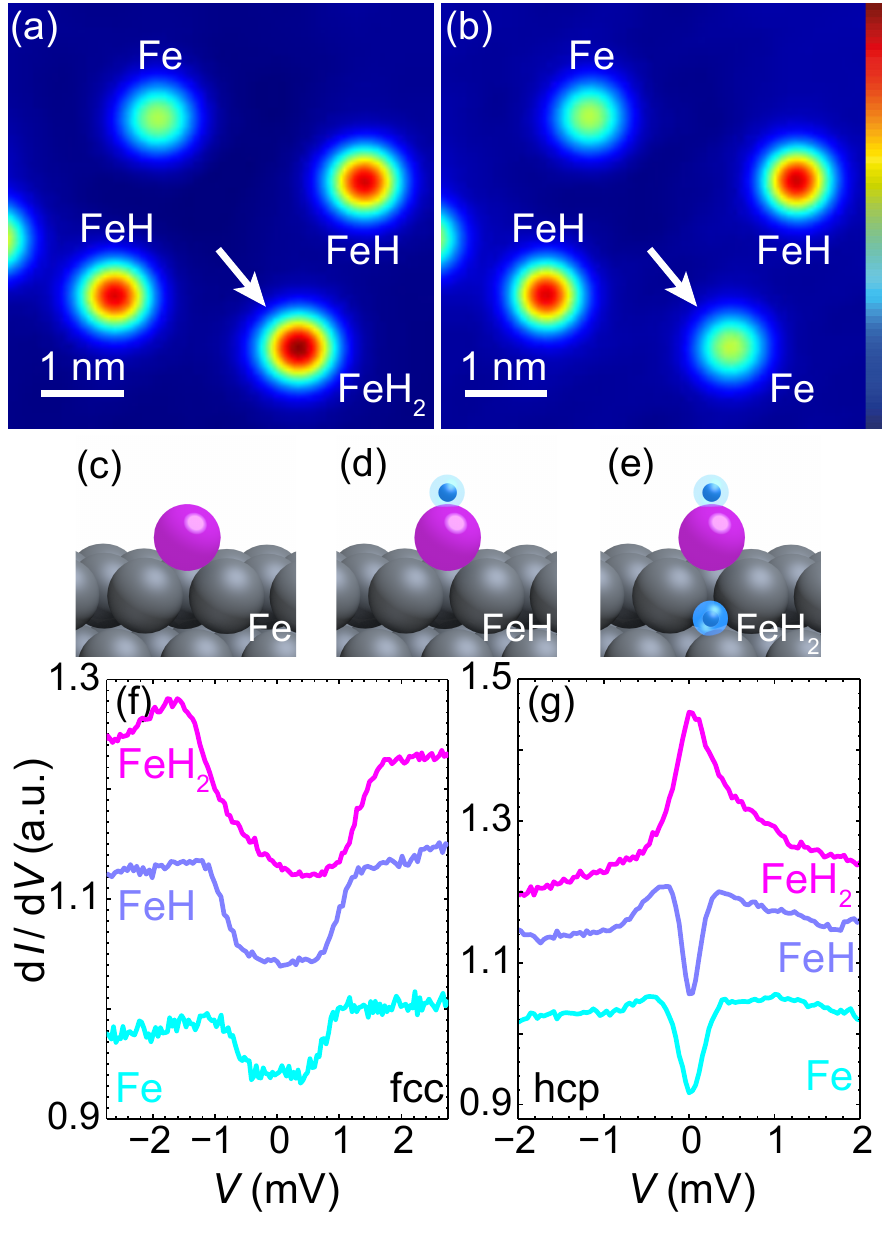}}
\caption{\label{fig2} \jw{{\bf Tuning the magnetism of Fe impurities.}} (a,b) STM images of \jw{clean} Fe and Fe-hydrogen \jw{complexes adsorbed} on the Pt(111) surface \jw{(heights of Fe, FeH, and FeH$_{2}$: $110$ pm, $190$ pm, and $200$ pm)}. \jw{The} arrow indicates an FeH$_{2}$ (a) before, and (b) after controlled dehydrogenation with a voltage pulse ($V_{\rm{pulse}} = 500$ mV, $V_{\rm{S}} = -100$ mV, $I_{\rm{t}} = 1$ nA). \jw{(c, d, e) Side views of DFT-calculated positions of Fe (pink spheres), hydrogen (blue spheres), and Pt atoms of the substrate (grey spheres) for the three different types of complexes. (f,g) ISTS of all six Fe-hydrogen complexes with increasing hydrogen coverage from bottom to top. Spectra} are vertically offset for visual clarity. (stabilization parameters: $|V_{\rm{S}}| = 5-10$ mV, $I_{\rm{t}} = 3$ nA, $V_{\rm{mod}}= 0.04$ mV).}
\end{figure}

The \jw{$3d$ transition metal impurities} we investigate are Fe atoms strongly coupled to a metallic platinum(111) surface by adsorption (Fig.~\ref{fig2}). Depending on the hollow site of the Pt(111) surface to which the Fe adatom is adsorbed, called \fcc and \hcp, the Fe adatom exhibits an out-of-plane easy axis, or an easy-plane anisotropy, respectively, with a considerable anisotropy strength on the order of a milli electronvolt~\cite{Khajetoorians2013PRL}. Accordingly, the inelastic scanning tunneling spectra (ISTS)~\cite{Heinrich2004} show a characteristic shape with symmetric steps around zero bias voltage $V$ stemming from the spin excitations of the local spin by tunneling electrons (Fig.~\ref{fig2}(f,g)). This qualitative change of the magnetic anisotropy between \fcc and \hcp originates from the site-dependent interplay of the magnetic anisotropy of the adatom and of the giant cloud of polarized Pt beneath the Fe adatom~\cite{Khajetoorians2013PRL}. The anisotropy can be controllably switched between out-of-plane and easy-plane, by moving the atom between fcc and hcp, and vice versa, using the tip of the STM as a fabrication tool~\cite{Khajetoorians2013PRL}. After long term exposure of the sample to residual H$_{2}$ gas~\cite{supp}, two new adsorbates on each of the two adsorption sites, which appear with an increased height, can be observed (Fig.~\ref{fig2}(a)). There is further experimental evidence and indication from our density-functional theory (DFT) calculations~\cite{supp}, that the additional adsorbates are singly hydrogenated Fe atoms containing one atomic hydrogen adsorbed to the top of \fcc or \hcp, labelled \Feonefcc and \Feonehcp, and doubly-hydrogenated Fe atoms where the second atomic hydrogen occupies the tetrahedral pore inside the first Pt layer underneath the \Feonefcc or \Feonehcp, labelled \Fetwofcc and \Fetwohcp (Fig.~\ref{fig2}(c,d,e)). Independent of the adsorption site, both FeH and FeH$_{2}$ can be controllably dehydrogenated by placing the STM tip over the impurity and applying a voltage pulse, thus recovering a clean Fe adatom \jw{without changing the adsorption site} (shown in Fig.~\ref{fig2}(a,b) for FeH$_{2}$).

ISTS of each of the six different impurity types (Fig.~\ref{fig2}(f,g)) illustrates that both the adsorption site and the degree of hydrogenation have a strong effect on the magnetism of the impurity. This is revealed by the different step voltages $V$ indicating different inelastic excitation energies $(E=eV)$, spectral lineshapes, and step heights indicating different excitation intensities ($\mathscr{I}$) of each impurity. For fcc impurities (Fig.~\ref{fig2}(f)), full hydrogenation increases the zero-field inelastic excitation energy $E(B_{z}=0)$ and $\mathscr{I^{\rm{fcc}}}$ almost by a factor of two. The most dramatic effect is observed for the hcp impurities (Fig.~\ref{fig2}(g)). While $E(B_{z}=0)$ decreases to half of the value from \hcp to \Feonehcp, the spectrum completely transforms from step-shaped to peak-shaped for \Fetwohcp. \jw{As shown later, }this sharp resonance at the Fermi energy results from a considerable Kondo screening of the local \Fetwohcp spin by a cloud of conduction electrons of the Pt substrate~\cite{Madhavan1998}.

\subsection{Hund's impurity character of the iron-hydrogen complexes}
\begin{figure}[t!]
\centerline{\includegraphics[width = \columnwidth]{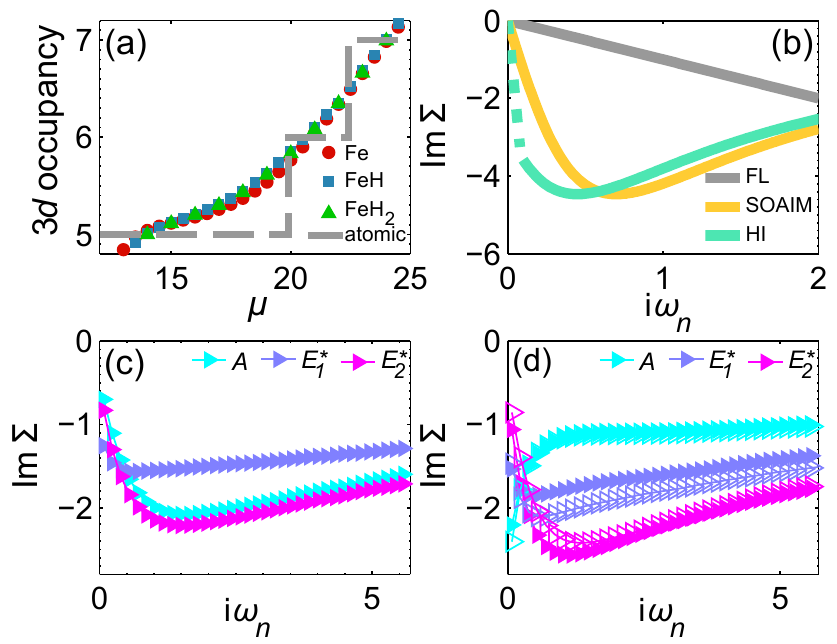}}
\caption{\label{fig3} \jw{{\bf Hund's impurity character of Fe-hydrogen complexes: orbital occupancy and scattering.} (a) QMC calculation of 3$d$ orbital occupancy $n$ versus applied chemical potential $\mu$ for the $\mathcal{A}$ limit and different Fe-hydrogen complexes at $T = 290$ K ($\beta = 1/k_{\rm B}T = 40 {\rm eV}^{-1}$). (b) Illustration of the behaviour of $\rm{Im} \Sigma(i \omega_n)$ for a Fermi liquid (FL), a strongly correlated FL within the single orbital Anderson impurity model (SOAIM) and a Hund's impurity ($\mathcal{HI}$). (c,d) QMC calculation of $\rm{Im} \Sigma(i \omega_n)$ at $T = 290$ K ($\beta = 40 {\rm eV}^{-1}$) for the orbitals of different symmetry {\it A} ($3z^2-r^2$), $E^*_1$ and $E^*_2$ of all six Fe-hydrogen complexes, i.e., \fcc and \hcp in (c), \Feonefcc and \Feonehcp closed triangles in (d), \Fetwofcc and \Fetwohcp open triangles in (d).}}
\end{figure}

In order to prove the $\mathcal{HI}$ character of these Fe-hydrogen complexes, we quantify the interactions illustrated in Fig.~\ref{fig1} and their effects on the 3$d$ orbitals of the complexes and the bath electrons by means of a five-orbital Anderson impurity model (AIM) \jw{solved numerically by quantum Monte Carlo (QMC) simulations (Fig.~\ref{fig3}) as well as exact diagonalization (ED)~\cite{supp}. All parameters of this model have been derived by means of DFT calculations~\cite{supp}. Of particular importance is the DFT-calculated {\it hybridization function} $\Delta(\epsilon)$ which is closely related to the strength of the hopping $V_{\rm{dk}}$~\cite{supp}. The degree of} proximity of an impurity to the \jw{$\mathcal{A}$} limit can be inferred from analyzing the \jw{orbital} occupation as a function of the chemical potential $n(\mu)$, which will yield a series of integer steps near the $\mathcal{A}$ limit (Fig.~\ref{fig3}(a)). When increasing the hybridization up to $\Delta \sim U_{\rm{Coulomb}}$, sizable charge fluctuations set in, resulting in a valency that is not well defined and the integer steps seen in the impurity occupation $n(\mu)$ smear out into a smoothly increasing function. This is indeed the case for all six Fe complexes (Fig.~\ref{fig3}(a)). Thus, all Fe-hydrogen complexes show metallic behavior and are far away from the $\mathcal{A}$ limit. The second requirement for the formation of a \jw{$\mathcal{HI}$} is the preservation of a sizeable impurity magnetic moment, which scatters the substrate electrons. Magnetic moment formation for all six impurities is demonstrated by our DFT calculations~\cite{supp}. The \jw{strength of} scattering of conduction electrons off these magnetic moments is inferred from the {\it Matsubara frequency} ($i\omega_n$) dependence of the imaginary part of the {\it self energy} $\rm{Im} \Sigma(i \omega_n)$ of the impurity electrons (Fig.~\ref{fig3}b).~\cite{supp} ~\cite{Georges2013,Werner2008,TW_Werner} The extrapolation of $\rm{Im} \Sigma$ to zero $i\omega_n$ is directly related to the strength of magnetic scattering. For a simple metal which can be modelled by a Fermi liquid (FL), $\Im \Sigma$ depends linearly on $i\omega_n$ and extrapolates to zero at sufficiently low temperatures. For strongly correlated FLs as realized, e.g., at intermediate hybridization in the single orbital AIM (SOAIM), $\left|\rm{Im} \Sigma\right|$ shows a maximum at intermediate $i \omega_n$, but still extrapolates to zero at low temperatures. For the $\mathcal{HI}$ case, $\rm{Im}\Sigma$ extrapolates to a nearly constant non-zero intercept in a certain intermediate temperature range~\cite{Georges2013,Werner2008,TW_Werner} and only at very low temperatures a strongly renormalized FL may emerge. Indeed, the combined DFT and QMC calculated~\cite{RMP_CTQMC} orbitally resolved self-energies of all six Fe impurities (Fig. \ref{fig3}(c,d)) reveal non-FL behavior, i.e., a finite scattering rate in the studied temperature range ($96 {\rm K} < T < 580 {\rm K}$), despite the apparently metallic behaviour in $n(\mu)$. This proves the $\mathcal{HI}$ character of all Fe-hydrogen complexes under investigation.

\subsection{Tuning magnetic anisotropy in the Hund's impurity}
\begin{figure}[t!]
\centerline{\includegraphics[width = 0.7\columnwidth]{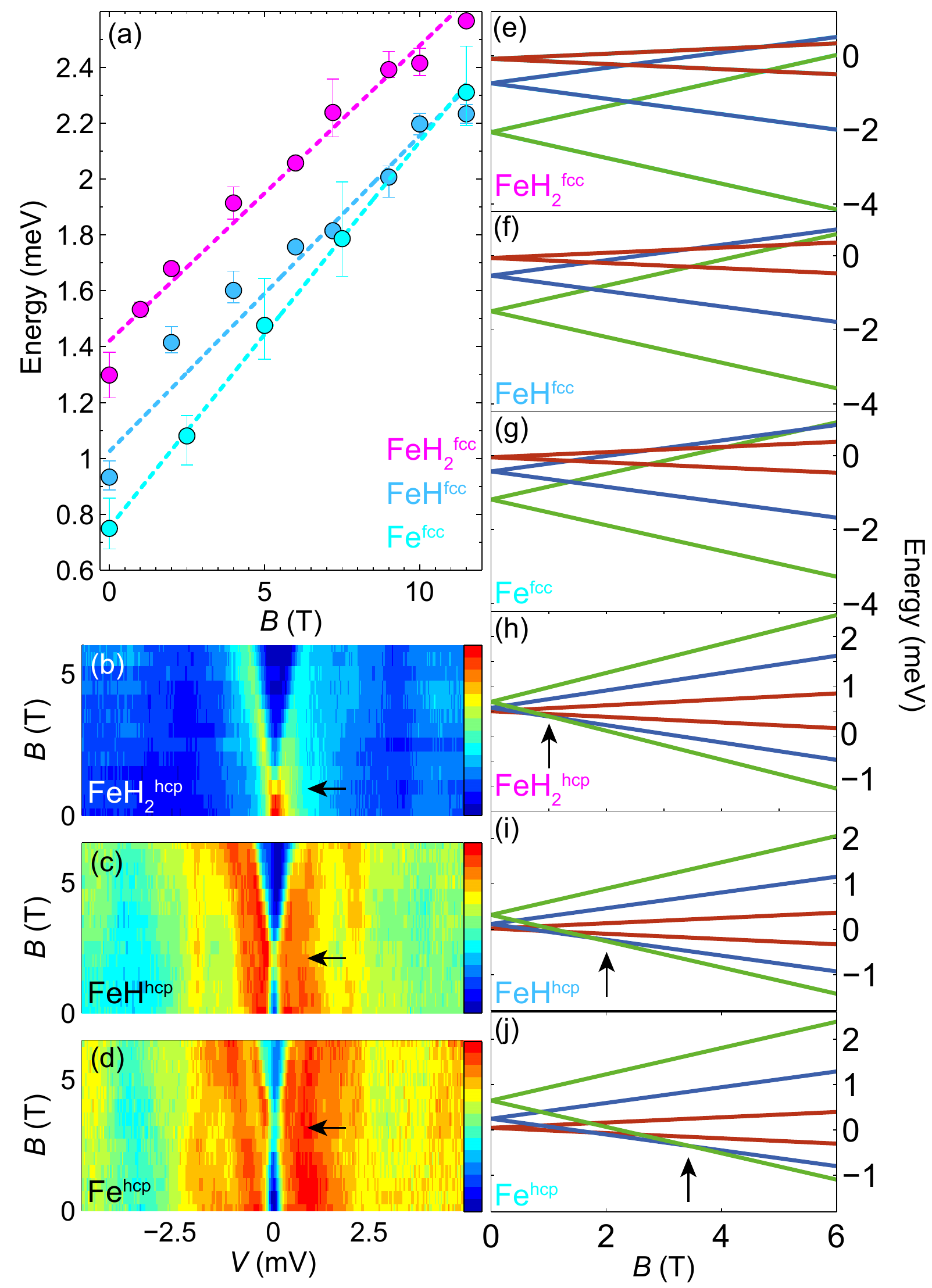}}
\caption{\label{fig4} {\bf Spin excitations of Hund's impurities:} (a) \jw{Dots: magnetic field dependency of the measured spin excitation energy }$E(B_{z})$ for \fcc, \Feonefcc, \Fetwofcc. \jw{Dashed lines: }linear fit to the measured $E(B_{z})$ \jw{resulting in the following parameters: $D = -0.33$ meV, $g = 1.83$ (\Fetwofcc); $D = -0.24$ meV, $g = 1.95$ (\Feonefcc); $D = -0.19$ meV, $g = 2.4$ (\fcc).} (b-d) Colorplot  representations of ISTS for \Fetwohcp, \Feonehcp, and \hcp (stabilization parameters: $|V_{\rm{S}}| = 5-10$ mV, $I_{\rm{t}} = 3$ nA, $V_{\rm{mod}}= 0.04-0.1$ mV). The arrows mark \jw{$B_{z}$ at which the spin excitation gap is minimal and where the linear Zeeman splitting starts.} (e-j) \jw{$B_{z}$ dependency of energetic eigenvalues of $\hat{\mathcal H}_{J}$ (red: $J_z=\pm1/2$, blue: $J_z=\pm3/2$, green: $J_z=\pm5/2$) resulting from fitting the effective spin model to the ISTS data: $D$ and $g$ for \fcc, \Feonefcc, \Fetwofcc given above; $D = 0.03$ meV, $g = 2.0$ (\Fetwohcp); $D = 0.05$ meV, $g = 1.9$ (\Feonehcp); $D = 0.1$ meV, $g = 2.0$ (\hcp). Arrows in (h-j) mark the level crossing of $J_z=+3/2$ and $J_z=+5/2$.}}
\end{figure}

In order to investigate the effect of hydrogenation on the magnetic anisotropy of the $\mathcal{HI}$s, ISTS was recorded as a function of applied magnetic field $B_{z}$ (Fig.~\ref{fig4}). Both \Feonefcc and \Fetwofcc show a linear increase in $E(B_{z})$ (Fig.~\ref{fig4}(a)), similar to clean \fcc \cite{Khajetoorians2013PRL}, revealing the out-of-plane easy axis of both impurity types. The $B_{z}$ dependence of ISTS on \Feonehcp (color-coded in Fig.~\ref{fig4}(c)) has a similar behaviour as that measured on clean \hcp (Fig.~\ref{fig4}(d)) although with a shifted position of the excitation energy minimum ($B_{z} \approx 2$ T, black arrow in Fig.\ref{fig4}(c)), revealing a diminished easy-plane anisotropy. \jw{The Kondo behaviour of \Fetwohcp (b) will be investigated in detail below, and we first focus on the other impurities.}


In order to extract the magnetic anisotropies of these impurities \jw{from the experimental data}, we use a model Hamiltonian which assumes an \jw{effective well defined spin $J$ of the local impurity moment in a crystal field with $C_{3v}$ symmetry:~\cite{Khajetoorians2013PRL, Chilian2011a}} $\hat{\mathcal H}_{J} = g\mu_{\rm{B}}B_{z}\hat{J}_z + D\hat{J}_{z}^{2}$ with the spin operator $\hat{J}_{z}$. The parameters of this effective spin Hamiltonian are the total angular momentum quantum number $J$, the magnetic anisotropy parameter $D$, where $D<0$ corresponds to out-of-plane easy axis and $D>0$ to easy-plane anisotropy, and the $g$ factor. We resort to the DFT-calculations~\cite{Khajetoorians2013PRL,supp} to estimate $J=5/2$ from the calculated magnetic moments $m$ of \fcc and \hcp~\cite{Khajetoorians2013PRL} which do not vary much upon hydrogenation for the various impurities~\cite{supp}. Please note, that the following discussion remains valid for other values of $J$ by adjusting $D$ accordingly. For \fcc, \Feonefcc, \Fetwofcc, \hcp, and \Feonehcp the experimental spectra are excellently reproduced over the whole $B_{z}$ range~\cite{supp}, using the parameters of $D$ and $g$ given in the caption of Fig.~\ref{fig4}. The corresponding energetic eigenvalues plotted in Fig.~\ref{fig4}(e-g,i,j) show a stepwise transition from strong out-of plane anistropy for \Fetwofcc, which decreases with less hydrogenation, to a strong easy-plane anisotropy for \hcp.

\subsection{\jw{Magnetic anisotropy and Kondo resonance of the \Fetwohcp Hund's impurity}}

\begin{figure*}[t!]
\centerline{\includegraphics[width = \columnwidth]{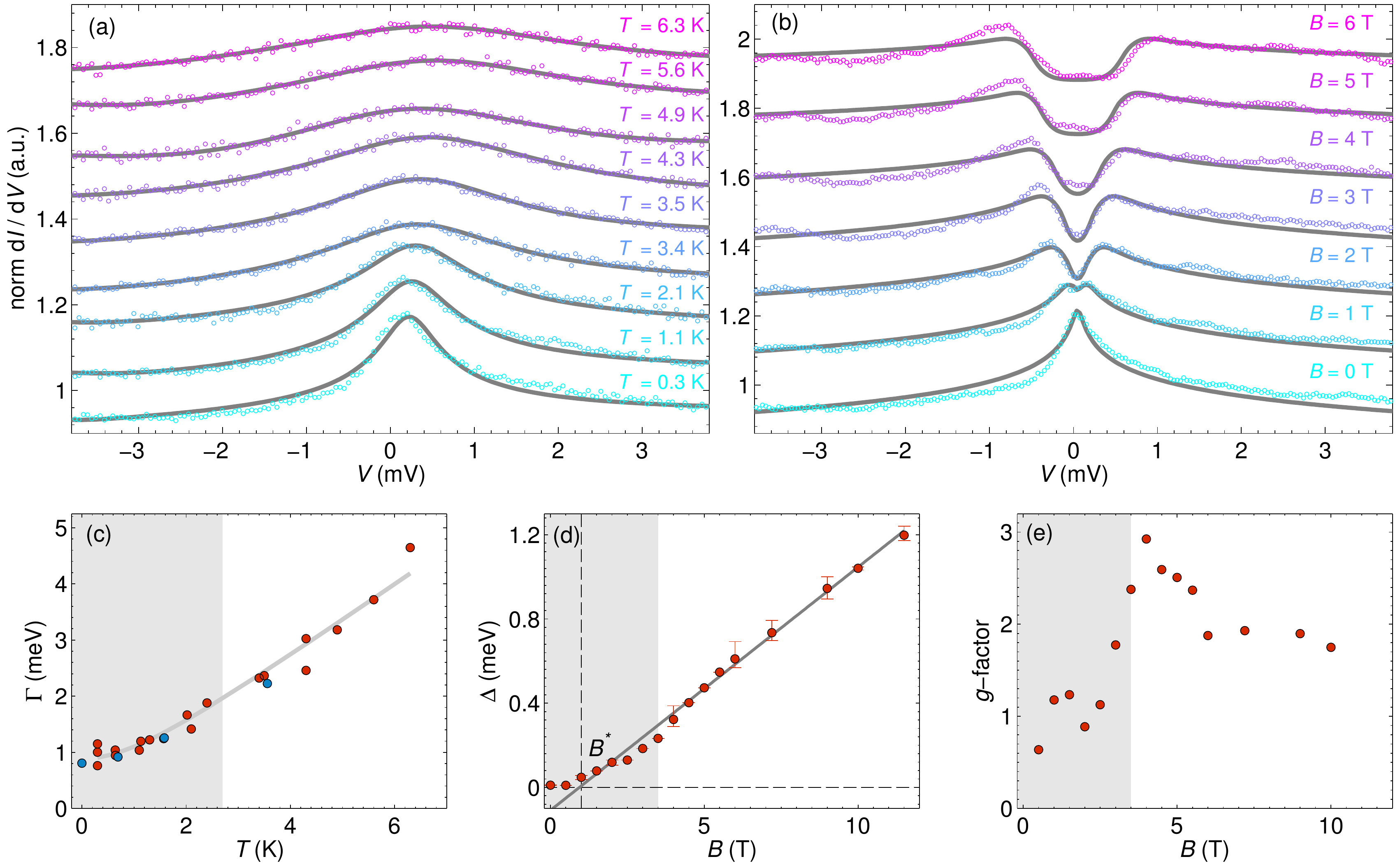}}
\caption{\label{fig5} {\bf Kondo screening of a Hund's impurity:} Colored dots in (a) and (b) show the temperature ($B_{z}= 0$ T) and magnetic field dependence ($T = 0.3$ K), respectively, of the Kondo resonance observed for \Fetwohcp (stabilization parameters: $|V_{\rm{S}}| = 6$ mV, $I_{\rm{t}} = 3$ nA, $V_{\rm{mod}}= 0.04$ mV). The gray lines indicate fits generated from a Frota function (a) and a spin-1/2 Anderson-Appelbaum model (b). \jw{(c) $\Gamma(T)$ from Frota fit (red dots) compared to NRG calculations \cite{Costi2000} for a spin-1/2 impurity in the strong coupling regime (blue dots). The grey line is a fit to a power law~\cite{supp}. Dots in (d) and (e) show $\Delta(B_{z})$ and $g(B_{z})$ from the Anderson-Appelbaum model fitting. Gray line in (d): linear fit to the data points outside of the shaded region ($g = 2$) intercepting the zero-$\Delta$ line at the characteristic magnetic field of $B^{*}\approx1$~T (dashed vertical line).}}
\end{figure*}

\jw{The $B_{z}$ dependence of ISTS measured on \Fetwohcp shows an evolution which is completely different from that of the other $\mathcal{HI}$s characterized by a temperature induced broadening (Fig.~\ref{fig5}(a)) and magnetic field induced splitting (Fig.~\ref{fig4}(b), Fig.~\ref{fig5}(b)) of a sharp peak which are the fingerprints of a Kondo resonance~\cite{Otte2008}. Despite the calculated strong hybridization and charge fluctuation of this $\mathcal{HI}$ the small value of the half-width half maximum (HWHM) of the resonance of $\Gamma\approx 1$~meV, and, consequently, strong sensitivity of the Kondo resonance to small values of $B_{z}$, indicates a Kondo temperature $T_{\rm{K}}$ that is very low as compared to other $3d$ transition metal adatoms on other metallic surfaces~\cite{Madhavan1998}. Indeed, $J_{\rm{Hund}}$ is known to quench Kondo temperatures~\cite{Nevidomskyy2009} and a correspondingly low $T_{\rm K}$ is typical. For $B_{z}> 2$~T, the Kondo resonance transforms into a spin-excitation gap that linearly grows with $B_{z}$ with a similar field evolution as for \hcp and \Feonehcp (Fig.~\ref{fig4}(b-d)) indicating the gradual quenching of the Kondo screening. In order to estimate the magnetic anisotropy $D$ of this impurity, we therefore use the same effective spin Hamiltonian $\hat{\mathcal H}_{J}$ with $J = 5/2$ as before to fit the ISTS data at $B_{z}> 2$~T~\cite{supp}, and then extrapolate the eigenstates into the Kondo regime. The resulting anisotropy parameter $D = 0.03$ meV and level diagram (Fig.~\ref{fig4}(h)) reveal the lowest easy plane anisotropy of all six investigated $\mathcal{HI}$s.}

\jw{In order to investigate the Kondo screening of this $\mathcal{HI}$ in detail, we present a detailed analysis of the measured temperature- ($T$) (Fig.~\ref{fig5}(a)) and $B_{z}$- dependence (Fig.~\ref{fig5}(b)) of its spectral function. The HWHM $\Gamma(T)$ of the resonance (Fig.~\ref{fig5}(c)), which was extracted by fitting the experimental spectra to a Frota function~\cite{Pruser2011,Zhang2013,Zitko2011,supp} (grey curves in Fig.~\ref{fig5}(a)), reveals the characteristic rise with $T$~\cite{Nagaoka2002}. On the low temperature side $\Gamma(T)$ saturates at $\Gamma \approx 0.85$ mV below $T \approx 2$~K, which indicates that $T_{\rm{K}}$ is about this temperature~\cite{Nagaoka2002}. For comparison, $\Gamma(T)$ extracted from the numerically exact spin-$1/2$ numerical renormalization group (NRG) method using $\Gamma(T=0) = 0.85$ meV~\cite{Costi2000} (blue dots in Fig.~\ref{fig5}(c)) shows a very good agreement with the $\mathcal{HI}$ data exemplifying that a FL emerges for the $\mathcal{HI}$ at very low temperatures. Since the Kondo impurity is in the strong coupling limit at our lowest experimental temperature of $T = 0.3$~K, as indicated by the saturation of $\Gamma(T)$ at low temperatures, $\Gamma(0.3\rm{~K})$ can be used to extract $T_{\rm{K}}$ via Wilson's definition of the Kondo temperature, resulting in $T_{\rm K}^{\rm W} = 0.27 \Gamma/k_{\rm{B}}\approx 2.8$~K~\cite{Pruser2011,Zitko2011}.} Note, that, from the temperature dependence alone, we thus cannot discreminate between a $\mathcal{HI}$ and a spin-$1/2$ impurity.

In order to fully determine the underlying description of the Kondo screening of the $\mathcal{HI}$, we extracted the $B_{z}$ dependent splitting $\Delta$ of the Kondo resonance, by fitting the spectra to a spin-$1/2$ Anderson-Appelbaum model~\cite{Zhang2013,Wallis1974,supp} (grey lines in Fig.~\ref{fig5}(b)). The resulting $\Delta(B_{z})$ shown in Fig.~\ref{fig5}(d) reveals a linear behaviour for $B_{z}> 4$~T with a slope of $g = 1/\mu_{\rm B}\cdot\text{d}\Delta/\text{d}B_{z} \approx 2$ (Fig.~\ref{fig5}(e)). Most remarkably, there is a considerable non-linear behaviour for $B_{z}< 4$~T, i.e. in the Kondo regime (grey shaded region), with a non-constant $g$. Note, that this is also obvious from the horizontal axis intercept $B^{*}(\Delta=0)\approx1$~T of a line fitted to the high field data (grey line in Fig.~\ref{fig5}(d)). This value of $B^{*}$ coincides with the level crossing of $J_{z}=+5/2$ and $J_{z}=+3/2$  within the effective spin model (arrow in Fig.~\ref{fig4}(h)) providing additional evidence of the magnetic anisotropy of this $\mathcal{HI}$. Moreover, we observe a clear splitting in the Kondo resonance already at $B_{z} \approx 1$~T, in stark contrast to the prediction of the NRG calculation~\cite{Costi2000} of the spin-$1/2$ Kondo model, where a clear splitting occurs only above $B_{z}>0.5 \Gamma(T=0)/(g\mu_{\rm B}) \approx 7$~T. These deviations from numerically exact spin-$1/2$ Kondo theories call for a more realistic multi-orbital approach including magnetic anisotropy in order to correctly describe the Kondo screening in a $\mathcal{HI}$.

\subsection{Discussion}
\begin{figure}[t!]
\centerline{\includegraphics[width = \columnwidth]{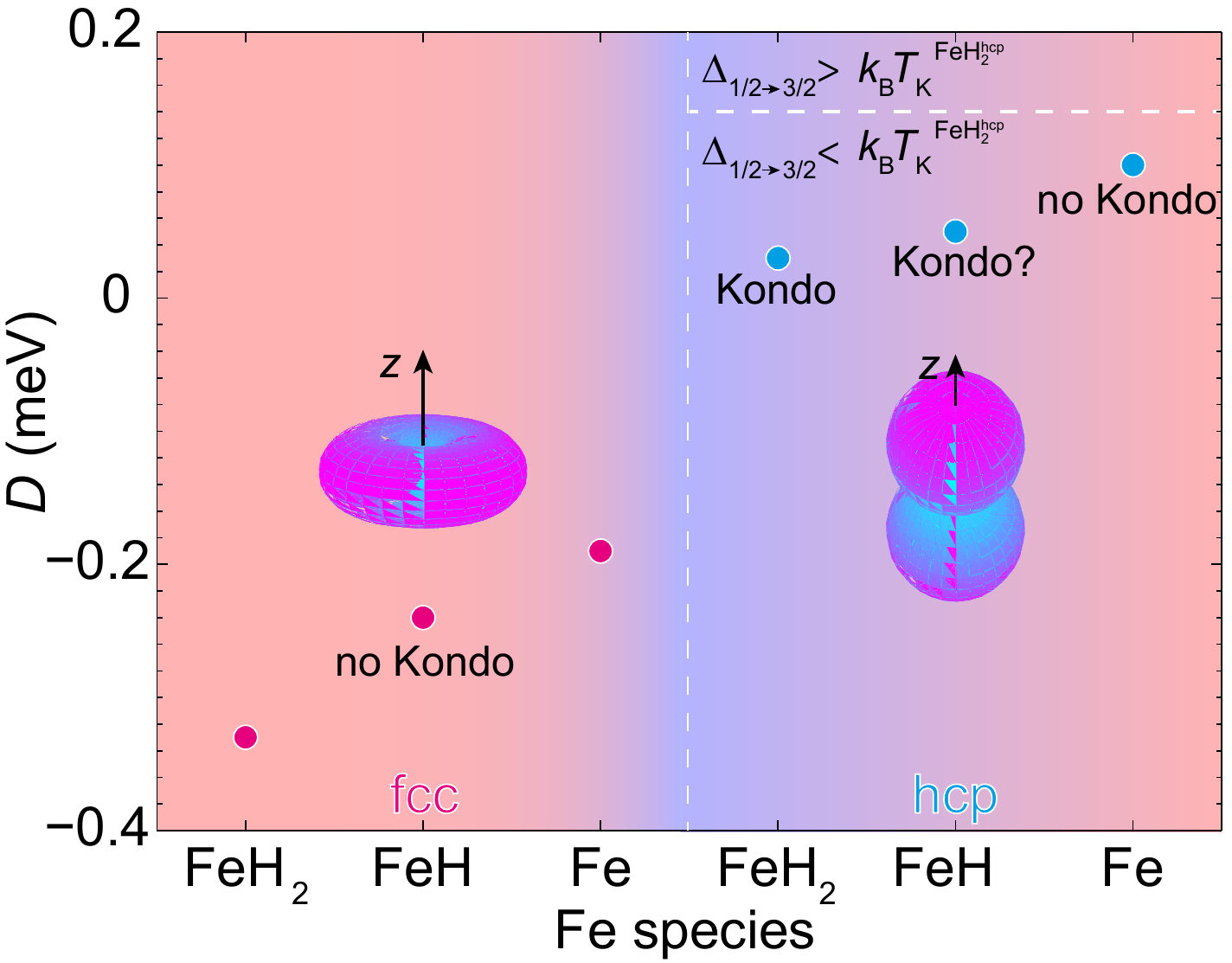}}
\caption{\label{fig6} {\bf Anisotropy and Kondo screening in the six Hund's impurities.} Value of the magnetic anisotropy parameter $D$ from effective spin model for each of the six Fe species. The tori indicate the corresponding surfaces of energy expectation values as a function of the orientation of the spin with respect to the surface normal ($z$) for the out-of plane easy axis ($D<0$) and easy plane ($D>0$) cases. Significant or negligible Kondo screening as revealed by the line shapes of ISTS of each impurity are indicated. The dashed horizontal line marks the $D$ value at which the splitting between the $J_{z}=\pm1/2$ and $ J_{z}=\pm3/2$ states gets larger than the energy scale of the \Fetwohcp Kondo screening.}
\end{figure}

The experimental \jw{and theoretical} results are summarized in Fig.~\ref{fig6}. Generally, the magnetic anisotropy of the $\mathcal{HI}$s can be tuned between out-of plane anisotropy or easy-plane anisotropy by a proper choice of the adsorption site between fcc or hcp, respectively, and fine-tuned in strength by additional adsorption of different amounts of hydrogen. As indicated by the different line shapes of the ISTS spectra (Fig.~\ref{fig2}(f,g)) only the \Fetwohcp $\mathcal{HI}$, which has the lowest magnetic anisotropy, reveals a strong Kondo screening. It has an energy scale of $k_{\rm B} T_{\rm K}$ that is four times larger than the magnetic anisotropy splitting $\Delta_{1/2\rightarrow3/2}$ between the $J_z=\pm 1/2$ and $J_z=\pm 3/2$ eigenstates of this $\mathcal{HI}$ within the effective spin model. Heuristically, within this effective spin model, the absence of a considerable Kondo screening for all fcc $\mathcal{HI}$s can be rationalized by the energetic-eigenvalue diagrams shown in Fig.~\ref{fig4} (e-g). For all those $\mathcal{HI}$s the Kondo effect is exponentially quenched due to their $J_z=\pm 5/2$ ground state which has a strong out-of plane anisotropy.~\cite{Otte2008} However, the situation is more complex for the hcp $\mathcal{HI}$s: All these $\mathcal{HI}$s possess a $J_z=\pm 1/2$ ground state, which in principle allows for strong Kondo screening even in the case of a large magnetic anisotropy,~\cite{Otte2008} although only \Fetwohcp of these $\mathcal{HI}$s shows a clear Kondo effect. There are two mechanisms which can explain this behavior. First, for $\Delta_{1/2\rightarrow3/2}$ smaller than the Kondo energy scale $k_{\rm B} T_{\rm K}$, the effective degeneracy of the impurity increases from two to four, which in turn leads to an increase of $k_{\rm B} T_{\rm K}$ as compared to the case of a $J_z=\pm 1/2$ ground state with larger magnetic anisotropy. Second, our DFT calculations show that the hybridization of the $d_{3z^2-r^2}$ orbital increases upon hydrogenation~\cite{supp}. Therefore, the corresponding QMC self-energies $|\Im \Sigma(i \omega_n)|$ also increase at small Matsubara frequencies $i \omega_n\to 0$ upon hydrogenation (Fig.~\ref{fig3} (c,d)), which signals strong quasi particle scattering and in turn facilitates Kondo screening. In other words, besides decreasing magnetic anisotropy, hydrogenation enhances the hybridization of the local $\mathcal{HI}$ spin with the substrate conduction electrons, leading to a trend towards a higher Kondo temperature of \Feonehcp and \Fetwohcp, with respect to \hcp. This trend is also in line with our ED simulations of~\Feonehcp and~\hcp cluster models~\cite{supp}, which show that hydrogenation can indeed control degeneracies of effective low energy spin degrees of freedom and that Kondo singlet ground states can emerge. Here, it is important to note, that we cannot rule out experimentally that \Feonehcp may also exhibit a weak Kondo screening which is split by the magnetic anisotropy at $B_{z}= 0$~T, as its ISTS reveals a faint peak structure on top of the spin excitation steps (Fig.~\ref{fig2} (g)).

\subsection{\jw{Conclusion}}


Our investigation shows that $\mathcal{HI}$s reveal a distinct two-faced nature of their magnetic moment, depending on the interplay between temperature, magnetic field, and magnetic anisotropy. At the one extreme, at high energies well above $k_{\rm B}T = k_{\rm B}T_{\rm{K}}$ and $\mu_{\rm{B}}B_{z} = k_{\rm{B}}T_{\rm{K}}$, the magnetic moment of a $\mathcal{HI}$ can be well described in terms of an effective spin model (Fig. \ref{fig4} (e-j)) despite the strong charge fluctuations, which are always present. The spin-excitations, which are solely derived from coupling to $B_{z}$ and the magnetic anisotropy, indeed do not rely on the precise value of the impurity spin explaining why the effective spin model works in these limits. At the other extreme, at low energies, FL behavior can reemerge due to Kondo screening of the impurity magnetic moment, which signals pure quantum behavior of the impurity magnetic degrees of freedom. Our experiments show that \hcp on Pt(111) realizes a case right at the border between these two extremes such that hydrogenation can tune the system from an emergent local moment (\hcp) to a correlated quantum state (\Fetwohcp) where a Kondo singlet forms. This system presents the smallest possible realization of what is referred to as a Hund's metal in bulk materials~\cite{Georges2013}. It provides an experimental and theoretical ground for understanding the complex physics from unconventional superconductors~\cite{Yin:2011,Werner2012} to transition metal oxides~\cite{Georges2013} and non Fermi liquids~\cite{Werner2008} in a bottom up way. Our work demonstrates that SOC largely controls electron correlations in $\mathcal{HI}$ systems and thus --- within the reasoning of dynamical mean field theory --- also in Hund's metals. The realization of topological order \cite{Balents2014} in materials governed by Hund's exchange and SOC remains to be explored. 

\section{Methods}
\jw{The experiments were performed using a scanning tunneling microscope (STM) facility~\cite{Wiebe2004} at temperatures between $T=0.3$~K and $T=6.3$~K and in magnetic fields, $B_{z}$, up to $12$~T perpendicular to the sample surface. Preparation of both tungsten STM tips and Fe adatoms on Pt(111) is described in ref.~\citenum{Khajetoorians2013PRL}. Adsorption of residual hydrogen from the ultra-high vacuum environment on the Pt surface and Fe adatoms typically appeared on the initially clean surface over the course of days once the cryostat remained cold for a prolonged period of time~\cite{Natterer2013, supp}. STM images were recorded in constant-current mode at a stabilization current $I_{\rm{t}}$ and with a bias voltage $V_{\rm{S}}$ applied to the sample. ISTS curves were taken by stabilizing the tip at ($I_{\rm{t}}$, $V_{\rm{S}}$), switching the feedback off and recording $\text{d}I/\text{d}V(V)$ via a lock-in technique with a modulation voltage $V_{\rm{mod}}$ (frequency $f = 4.1$~kHz) added to $V_{\rm{S}}$.}

A theoretical first principles description of the Fe and hydrogenated Fe impurity systems is based on DFT calculations done within the Vienna \textit{ab initio} simulation package (VASP) \cite{VASP} using projector augmented wave (PAW) \cite{PAW1,PAW2} potentials and generalized gradient approximation (GGA) \cite{PBE} to the exchange-correlation potential. We performed spin-polarized calculations to obtain adsorption geometries, magnetic moments, local density of states and hybridization functions~\cite{JPhys_2011_MK_TW,supp}. The latter were used to define Anderson Impurity Models, which are solved by means of exact diagonalization~\cite{supp} and hybridization expansion continuous time quantum Monte Carlo (CT-QMC) approaches~\cite{RMP_CTQMC} based on a segment algorithm~\cite{Werner_Troyer_Millis} and the Toolbox for Research on Interacting Quantum Systems (TRIQS)~\cite{triqs_web} solver.


\section{Acknowledgements}

We acknowledge fruitful discussions with the group of Samir Lounis. A. A. K., T. S., M. S., J. W. and R. W. acknowledge funding from SFB668-A1 and GrK1286 of the DFG. A.A.K. also acknowledges Project no.~KH324/1-1 from the Emmy-Noether-Program of the DFG. R.W. und J.W. additionally acknowledge funding from the ERC Advanced Grant "ASTONISH". M.V., T.O.W. and A. I. L. acknowledge support from FOR1346 of the DFG. M.V. and A. I. L. also acknowledge computer support NIC, Forschungzentrum J\"ulich, under project HHH14.

\bibliography{library}

\newpage

\newpage

\newpage


\newpage

\newpage

\newpage

\end{document}